\documentclass{class}
\title{\texorpdfstring{\begin{flushright}
\small LA-UR-24-24506 
\end{flushright}}{}%
Effect of helium bubbles on the mobility of edge dislocations in copper}

\author[1]{Minh Tam Hoang}
\author[2]{Nithin Mathew}
\author[3]{Daniel N. Blaschke}
\author[4]{Saryu Fensin}

\affil[1]{MST-8, Los Alamos National Laboratory, Los Alamos, NM, 87545, USA
and}
\affil[ ]{Department of Chemical Engineering and Materials Science, University of Minnesota Twin Cities, 151 Amundson Hall, 421 Washington Ave. SE, Minneapolis, MN
55455-0132, USA}
\affil[2]{T-1, Los Alamos National Laboratory, Los Alamos, NM, 87545, USA}
\affil[3]{XCP-5, Los Alamos National Laboratory, Los Alamos, NM, 87545, USA}
\affil[4]{MPA-CINT, Los Alamos National Laboratory, Los Alamos, NM, 87545, USA\vspace{0.3cm}}
\affil[ ]{\emph{email:} $^1$hoang370@umn.edu, $^2$mathewni@lanl.gov, $^3$dblaschke@lanl.gov, $^4$saryuj@lanl.gov\vspace{-0.8cm}}

\abstract{Helium bubbles can form in materials upon exposure to irradiation.  It is well known that the  presence of helium bubbles can cause changes in the mechanical behavior of materials.  To improve the lifetime of nuclear components, it is important to understand deformation mechanisms in helium-containing materials.  In this work, we investigate the interactions between edge dislocations and helium bubbles in copper using molecular dynamics (MD) simulations. We focus on the effect of helium bubble pressure (equivalently, the helium-to-vacancy ratio) on the obstacle strength of helium bubbles and their interaction with dislocations. Our simulations predict significant differences in the interaction mechanisms as a function of  helium bubble pressure. Specifically, bubbles with high internal pressure are found to exhibit weaker obstacle strength as compared to low-pressure bubbles of the same size due to the formation of super-jogs in the dislocation.
Activation energies and rate constants extracted from the MD data confirm this transition in mechanism and enable upscaling of these phenomena to higher length-scale models.}


\makeatletter
\AtBeginDocument{
	\hypersetup{
		pdfkeywords = {\@englishkeywords},
		pdftitle = {\@title},
		pdfauthor = {Minh Tam Hoang, Nithin Mathew, Daniel N. Blaschke, Saryu Fensin}
	}
}
\makeatother

\graphicspath{{./}{./Images/}}
\begin{document}
\setlength{\parskip}{0pt}
\raggedbottom
\bstctlcite{IEEEexample:BSTcontrol}
\maketitle

\section{Introduction}

In nuclear environments, radiation-induced defects and transmutation elements \cite{Choppin_alpha_decay} alter material behavior \cite{challenges,challenges_2} due to changes in phase stability \cite{phase_instability,RUSSELL1984229},  void swelling \cite{void_damage,void_damage_2}, and segregation \cite{TUOMISTO202044,segregration_2},
as well as the formation of various microstructural defects 
\cite{dislocation_loop,dislocation_loop2,void_formed,stacking,stacking_2,DEMKOWICZ2020152118,PhysRevLett.124.167401}.
The low solubility and diffusion of He atoms, coupled with their binding to vacancies can lead to the formation of He bubbles with varying internal pressures \cite{LI2022117656, WANG2020152051, RAO2022153613}.
Interactions between H isotopes and He have also shown to have an effect on irradiation-induced void swelling \cite{MUKOUDA2000302} and mobility of He bubbles \cite{LI20176902}. A detailed review of these, in fusion-relevant materials, can be found in ref. \cite{Hammond_2017}.
Additionally, He atoms can also segregate to other defects such as dislocations and grain boundaries \cite{martinez_acta_mat,martinez_nucl_fusion}, affecting their mobility. Thus, He atoms and bubbles can have a significant impact on  the overall microstructural evolution \cite{Eldrup} and consequently, result in changes in work hardening and ductility, ultimately leading to changes in material strength. Therefore, it is important to understand the mechanisms involved in this process as a  function of He bubble pressure and morphology.

There has been previous research on the effect of He bubbles on properties of materials \cite{microstructure_yamamato, microstructure_maziasz, microstructure_stoller, stoller_1990, Ghoniem, Adams, tungsten, iron_1, Marian, Wirth_2004,Cui:2023,Sills:2024}.
Specifically, Sch{\"a}ublin and Haghighat investigated the effects of He on hardening mechanisms in iron using molecular dynamics (MD) simulations \cite{iron_1, iron_2}.
This behavior has also been observed in Copper implanted with He \cite{LEAR2023118987}.
In particular, increase in yield strength and hardness was observed in samples tested at slow strain rates (0.001 - 0.05 Hz).
A related recent study in \cite{Jian:2022} also found that He bubbles can act as obstacles to dislocations leading to strengthening of materials.
Recent studies found that He bubbles in Cu can alter the plastic deformation in Cu by facilitating dislocation storage and interfering with dislocation motion \cite{Ding,situ_cu,neogi,ding_2,Shargh:2023}.
An atomistic study by Neogi \textit{et al.} \cite{neogi} demonstrated the relationship between the inter-bubble distance and the plastic behavior of single crystalline Cu nanopillars (NP).
While these bodies of work separately highlighted the importance of He-bubble features, such as size, spacing and internal pressure on the amount and the mechanisms for hardening, there is no systematic study that investigated each of these variables.
Previous research has shown that the equilibrium helium-to-vacancy ratio ranges between 0.4 to 0.7 for bubbles with diameters of 2--5 nm at room temperature and slightly decreases as temperature increases \cite{helium_eq, electron_microscopy}. Recent experimental work from members of our team also shows presence of He bubbles with an average diameter $\approx 2nm$ in He-implanted Cu \cite{LEAR2023118987}.
Another study in Fe \cite{osetskystoller} showed that He bubbles with helium-to-vacancy ratios of He:Vac $\leq 1:1$ exhibit the same dislocation-bubble interaction mechanism as observed in voids \cite{iron_3}.
In contrast, higher pressure He bubbles (2He:1Vac) lead to the emission of interstitial atoms that bind to dislocations and form super-jogs. Additionally, higher pressure He bubbles do not shear upon dislocation interaction, as is the case for voids and bubbles with low internal pressures, leading to an approximately $\sim40\%$ decrease in obstacle strength.

According to these studies, both theoretical and experimental works on the effect of He bubbles in metallic materials can be divided into several categories: bubble nucleation, bubble growth/coalescence, bubble-dislocation interaction, and bubble rupture. As for the third category, even though there has been a significant amount of work conducted on BCC Fe \cite{iron_1,iron_2,Qiu:2023} and BCC W \cite{tungsten,Hammond:2017b,Sandoval:2018}, there is still a lack of literature for FCC Cu. Motivated by the discrepancy in the mechanical behaviour of He bubbles in previous studies, we conduct an investigation on the interaction between He bubbles and moving edge dislocations in FCC Cu at an atomic level.

In this work we use molecular dynamics (MD) to address the following two central questions:
\begin{enumerate}
\item How does the obstacle strength of bubbles depend on their size and He content and how do these properties affect dislocation bubble interaction?
\item How can we rationalize He effects on dislocation-bubble interaction (hardening mechanism) based on quantitative factors including the pressure of He bubbles (He:Vac ratio), temperature, bubble size, and resolved shear stress?
\end{enumerate}

Several mechanisms, such as interactions between He bubbles and defects such as dislocations and grain boundaries, occur on length and time-scales that are still prohibitive for first principle (\textit{ab initio}) methods. Consequently, the development of predictive models about the effects of He on material strength requires targeted atomistic simulations using empirical/semi-empirical potentials. In fact, molecular dynamics (MD) simulations have been widely utilized to study the interaction between He bubbles and mobile defects such as dislocations as well as the impact of helium atoms and/or helium-vacancy clusters on materials properties within the scope of microstructural evolution.

\section{Methodology}

\begin{figure}[htb]
\includegraphics[width=\textwidth,trim=0 7.9cm 0 6.5cm, clip]{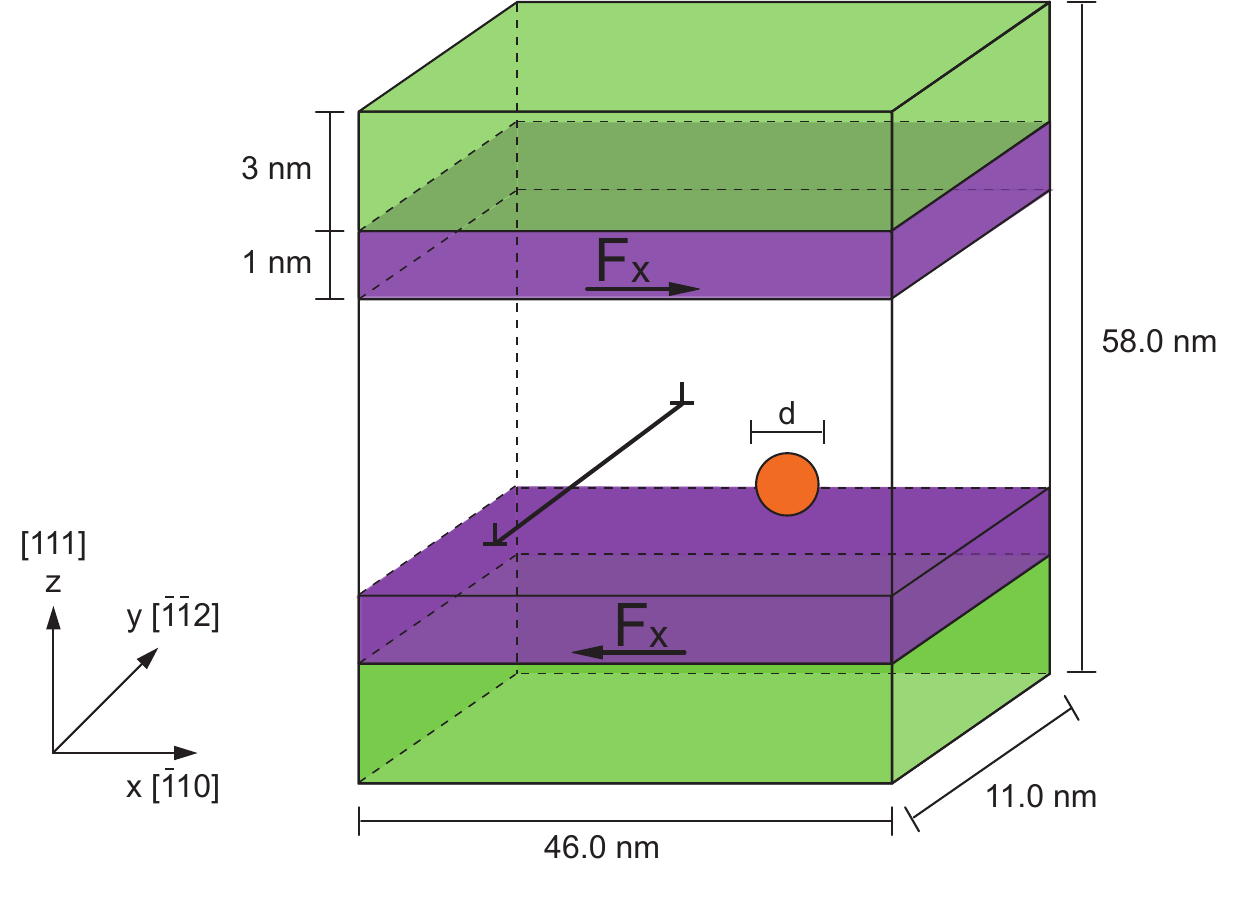}    
\caption{Schematic representation of the simulation box. The purple and green regions represent the region for applying the force and vacuum, respectively. The symbol $\perp$ specifies the edge dislocation, and the He bubble with diameter d stays on its glide plane. A force $\mathrm{F_x}$ is applied along the positive x direction on the top region and along the negative x direction on the bottom region.} 
\label{fig:simulation_box}
\end{figure}

In this work, MD simulations were utilized to study the interaction of an edge dislocation with a He bubble. All simulations were performed using the Large-scale Atomic / Molecular Massively Parallel Simulator (LAMMPS) package \cite{lammps} with the Embedded Atom Method (EAM) inter-atomic potential for Cu-He  \cite{CuNb_potential, Kashinath_2011}. Formation energies of He defects and Cu-He interaction forces predicted by this potential were found to be in good agreement with DFT as shown in \cite{Kashinath_2011}. A schematic of our simulation setup is shown in Fig. \ref{fig:simulation_box}. The initial configuration of single crystal Cu contained an edge dislocation with Burgers vector \textbf{b} = $\frac{1}{2}[\overline{1} 1 0]$ and was oriented along $x = [\overline{1} 1 0]$, $y = [\overline{1} \overline{1} 2]$, and $z = [1 1 1]$. The edge dislocation was created using the Periodic Array of Dislocations (PAD) method \cite{osetsky_bacon} as implemented in ATOMSK \cite{HIREL2015212}. The dimensions of the simulation box were approximately 460\,\AA, 110\,\AA, and 580\,{\AA} in the x,y, and z directions, respectively, with $\sim2.28$ million atoms. These simulation dimensions were chosen based on guidelines developed by Szajewski and Curtin \cite{Szajewski_2015} to minimize spurious image forces during dislocation-obstacle interaction.  Size effects were checked by investigating the effect of the simulation cell size (ranging from $\sim300,000$ to  $\sim2.28$ million atoms), on the Critical Resolved Shear Stress (CRSS) at 0 K to bypass helium bubbles with the ratio 3 He:1 Vac and considering the following total numbers of He$_n$:Vac$_m$, i.e.: He$_3$:Vac$_1$, He$_{36}$:Vac$_{12}$, and He$_{135}$:Vac$_{45}$.
These size-effect studies were performed keeping the aspect ratio and the obstacle spacing along the dislocation line direction constant at values of $\sim0.8$ and $\sim11$ nm, respectively. The CRSS values converged at system sizes of $\sim1.7$ - 2.3 million atoms. Periodic boundary conditions were applied along the x and y-axes, whereas the z-axis was non-periodic with a vacuum region of approximately 3 nm thickness. The conjugate gradient method was used to minimize the initial structure. The perfect dislocation \textbf{b} in the glide plane dissociated into two Shockley partials with Burgers vectors of $\frac{1}{6}[\overline{1} 2 \overline{1}]$ and $\frac{1}{6}[\overline{2} 1 1]$ with an enclosed stacking fault during the energy minimization process. 

Helium bubbles with radii of 5 {\AA} and 10 {\AA} were considered and modeled at different helium-to-vacancy ratios (He:Vac) ranging from 1:1 to 3:1. These specific He:Vac
ratios were chosen to demonstrate changes in interaction mechanisms due to over-pressurization. Ratios beyond 3:1 were avoided to prevent bubble instability, as previous work has reported the formation of Frenkel pairs in 2 nm radius bubbles with He:Vac $\mathrm{>}$ 3, indicating unstable conditions \cite{haghighat2009state}. Additionally, experimental and molecular dynamics studies typically report bubble diameters of less than 6 nm \cite{Ding, ding_2, Jian:2022, baskes1986recent}. Therefore, we selected these specific He:Vac ratios and sizes to ensure they are within the range of previous work, allowing us to compare our findings. The He bubble was introduced in the simulation box as follows: A sphere was first defined inside the simulation box with the center on the glide plane; copper atoms were then removed and replaced with different numbers of He atoms, followed by energy minimization. The bubble was placed $\sim11.5$ nm away from the dislocation to minimize interactions between the two defects prior to the gliding of the dislocation. Regions with a width of 1 nm were defined at the top and bottom of the simulation box, to apply a force, $F_x$, to drive the dislocation. The applied forces were anti-parallel in the glide plane with directions of $[\overline{1} 1 0]$ and $[1 \overline{1}  0]$. The magnitude of the forces, added to each atom, were calculated from the resolved shear stress as $\sigma_{xz} \cdot A_{xy}/N_{upper}$ (upper region) and $-\sigma_{xz} \cdot A_{xy}/N_{lower}$ (lower region) where $A_{xy}$ is the area of the x-y plane of the simulation box and $N_{upper}$ and $N_{lower}$ are the number of atoms in the upper and lower regions. The molecular dynamics simulations were performed in the micro-canonical ensemble (time step = 1 fs) in combination with the Langevin thermostat (damping =1 ps) for 0.1 ns at zero stress and for 1--4 ns under constant applied resolved shear stress. The simulations were conducted at a range of temperatures from 300 K to 400 K. The pressure in the bubble was calculated using the per-atom stress tensor, as shown in equation \ref{press_equation}:
\begin{equation}
    \label{press_equation}
    p = -\frac{1}{3} \frac{\sum_{i=1}^{n} s^i_{jj}}{\sum_{i=1}^{n} v^i}. 
\end{equation}
Here $p$ is the pressure, $s^{i}_{jj}$ is the $jj$ component of the per-atom stress tensor for the $i$th He atom, $n$ is the total number of He atoms, and $v^i$ is the Voronoi volume \cite{rycroft2009voro++} of the $i$th He atom. Summation over repeated indices is implied.
In order to quantify the obstacle strength of He bubbles, we calculated the energy barrier for the dislocation to overcome the bubble. The simulations were performed under a range of applied shear stresses and temperatures for each type of He bubble, to obtain the activation energies and pre-factors for this process. These were estimated using the Arrhenius equation for the temperature dependence of the rate at which dislocations overcame He bubbles. This rate ($k$) was determined as $ k= 1.0/(t_f-t_i)$, where $t_i$ ($t_f$)  is the simulation time at which the leading (trailing) partial is 4 nm away from the center of mass (COM) of the He bubble. Once the rates corresponding to different $T$ and $\sigma$ were obtained, $E_a$ and $V$ were extracted using the following equation:
\begin{equation}
    \label{rate_equation}
    k = k_o \exp{\left(\frac{-E_a(\sigma)}{k_B T}\right)}, 
\end{equation}
where $k_o$ is the pre-factor, $E_a$ is the activation energy, $\sigma$ is the resolved shear stress, $T$ is the temperature, and $k_B$ is Boltzmann's constant.
When the energy barriers are sufficiently low for thermal energy (${k_B T}$) to be significant, thermal fluctuations can assist dislocations to glide.

Common neighbor analysis (CNA) \cite{CNA} and the centrosymmetry parameter \cite{centrosymmetry} were used to analyze the atomistic configuration during the interaction between the dislocation and He bubble. The dislocation extraction algorithm (DXA) \cite{DXA} was utilized to extract the dislocation types. All of these analyses were performed using the Open Visualization Tool OVITO \cite{ovito}.

\section{Results and Discussion}
\subsection{Interaction of moving edge dislocations with 5\AA-radius He bubbles}

\begin{figure*}[!htbp]
\centering
\includegraphics[width=\textwidth]{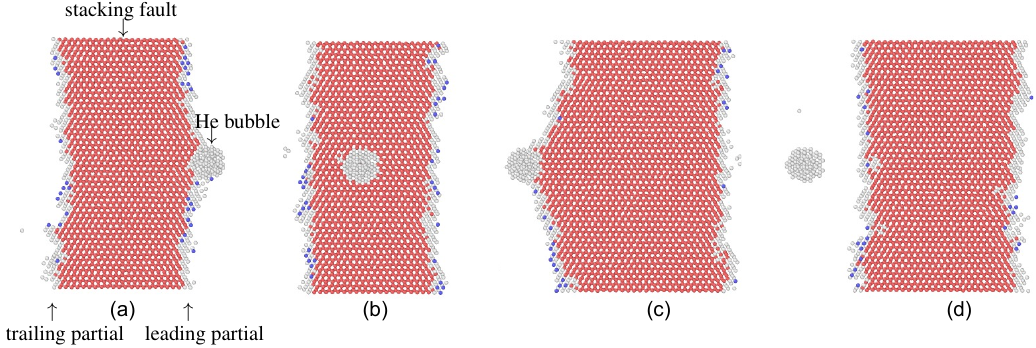}
\caption{Top-down view onto the glide plane showing the interaction between an edge dislocation moving here from left to right and a 5-{\AA} radius He bubble in FCC Cu at an applied shear stress of 85 MPa. There are four main stages of  interaction: a) the leading partial dislocation attracted by the He bubble, b) dislocation gliding through the bubble c) the trailing dislocation bowing before the release from the bubble, d) dislocation moving away from the bubble after the interaction.
All atoms are colored according to their crystal structure, i.e. body-centered cubic (BCC) atoms are blue, hexagonal close-packed (HCP) atoms are red, and disordered atoms white.
All sub-figures are snapshots in time capturing the various main stages and we have highlighted the main features with annotations in the first time snapshot for the reader's convenience.} 
\label{fig:1X_5A_300K_85MPa_s1}
\end{figure*}
Fig.~\ref{fig:1X_5A_300K_85MPa_s1} illustrates atomic configurations during the dislocation motion in the case of a 5 {\AA} He bubble with 1He:1Vac at an applied stress of 85 MPa at  300 K. There are four characteristic stages observed in the mechanism. Stage 1 (Fig.~\ref{fig:1X_5A_300K_85MPa_s1}.a) shows the initial contact of the leading partial with the He bubble. The interaction continues as the leading partial dislocation is released from the bubble and the trailing partial glides towards the bubble (Fig.~\ref{fig:1X_5A_300K_85MPa_s1}.b). Stage 3 (Fig.~\ref{fig:1X_5A_300K_85MPa_s1}.c) shows bowing of the trailing partial. The extent of bowing can be characterized by an angle between the dislocation segments and the bubble as discussed in  \cite{hullbacon}.
The mechanism described above ends with the release of the dislocation from the bubble and the restoration of the dislocation line (Fig.~\ref{fig:1X_5A_300K_85MPa_s1}.d). Cross slip, jog formation, or interstitial loops are not observed indicating that the bubble is sheared by the dislocation. This mechanism is observed for He bubbles with He:Vac $\leq$ 2:1. However, there exist minor differences between dislocations interacting with He bubbles of varying pressure. As the He:Vac ratio increases from 1:1 to 2:1, the critical bowing angle decreases and more dislocation segments of screw character formed from the trailing partial dislocation. Previous work has shown that the angle between dislocation segments attached to the obstacle is inversely proportional to the obstacle strength \cite{hullbacon}, and therefore, a smaller angle for 2He:1Vac suggests that it is a stronger obstacle compared to 1He:1Vac bubble.

\begin{figure*}[b]
\centering
\includegraphics[width=\textwidth]{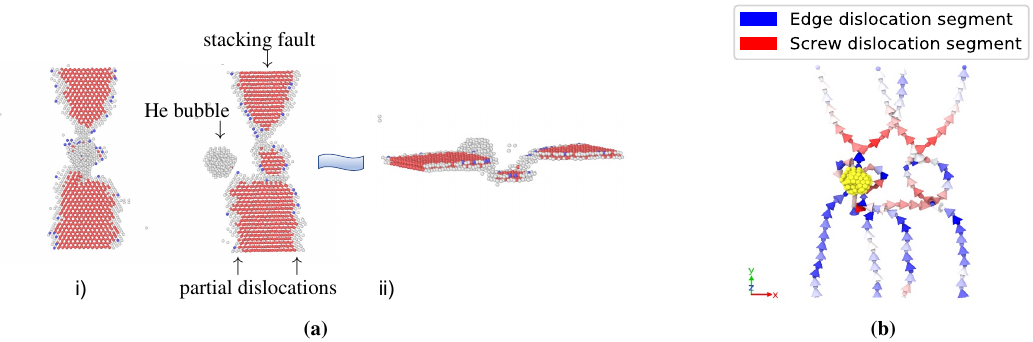}
\caption{(a) Atomic snapshots representing i) the intersection of two partial dislocations and ii) the formation of a super-jog;
Atoms are colored according to their crystal structure, i.e. body-centered cubic (BCC) atoms are blue, hexagonal close-packed (HCP) atoms are red, and disordered atoms white.
(b) Configurations of dislocation lines that correspond to two stages described in (a). }%
\label{fig:compare_dislocation_line_3x}%
\end{figure*}

To demonstrate major differences in the interaction mechanism between different He:Vac ratios, we present the interaction between dislocations and He bubbles with 3He:1Vac in Fig.~\ref{fig:compare_dislocation_line_3x}. The variations in the mechanism are shown explicitly prior to and after the release of the dislocation from the bubble. It can be seen that the stacking fault is constricted, during the interaction of the dislocation with the bubble, resulting in intersection of the partials. This is starkly different from the mechanism depicted in Fig.~\ref{fig:1X_5A_300K_85MPa_s1}, where the partials shear the bubble and there is no constriction.
The larger pressure of the 3He:1Vac bubble
distorts the neighboring matrix and consequently, promotes an emission of self-interstitial atoms (SIAs) during the interaction. These SIAs are absorbed by the dislocation leading to the formation of a super-jog. This SIA-enabled super-jog formation enables the dislocation to bypass the bubble. 

\begin{figure}[t]
\includegraphics[width=\textwidth]{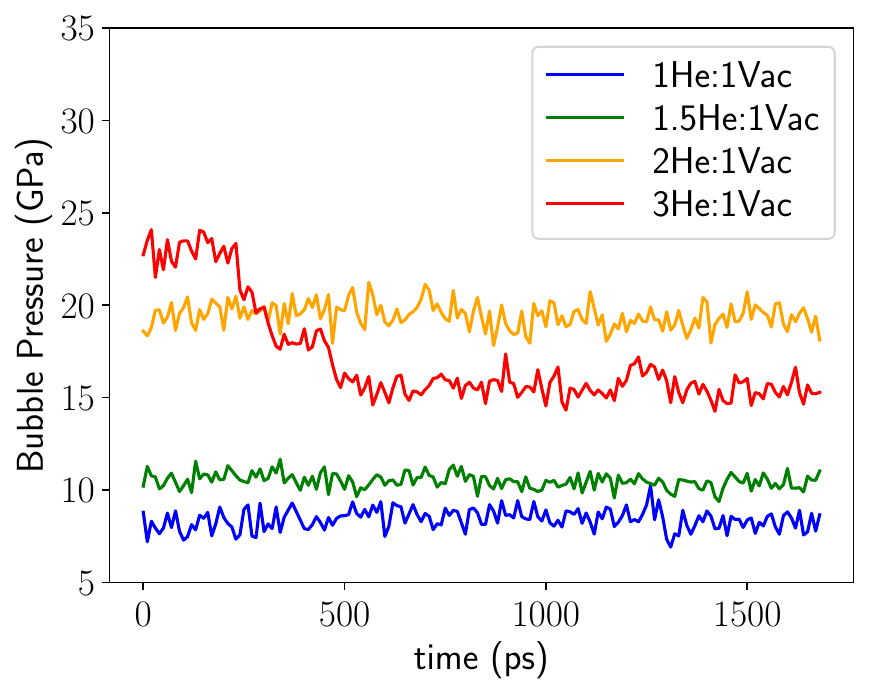}    
\caption{Changes in the bubble pressure during the interaction between dislocations and 5\AA-radius bubbles with 1He:1Vac, 1.5He:1Vac, 2He:1Vac, 3He:1Vac.} 
\label{fig:BubblePressure}
\end{figure}

The change in the bypass mechanism is evident in the variation of the bubble pressure during the interaction with the dislocation shown in Fig.~\ref{fig:BubblePressure}. No significant change in the pressure is predicted during the interaction for bubbles with He:Vac=1, 1.5, and 2. In contrast, there is a drop in the pressure for He:Vac=3, which indicates emission of Frenkel pairs from the vicinity of the bubble. The vacancies in the Frenkel pairs are absorbed by the bubble, which results in the drop in pressure.
We note that the mechanism at He:Vac=2.5 was similar to He:Vac=2 and therefore, this case was not investigated in detail.

\begin{figure}[!htbp]
\includegraphics[width=\textwidth,trim=0 0.2cm 0 0.2cm, clip]{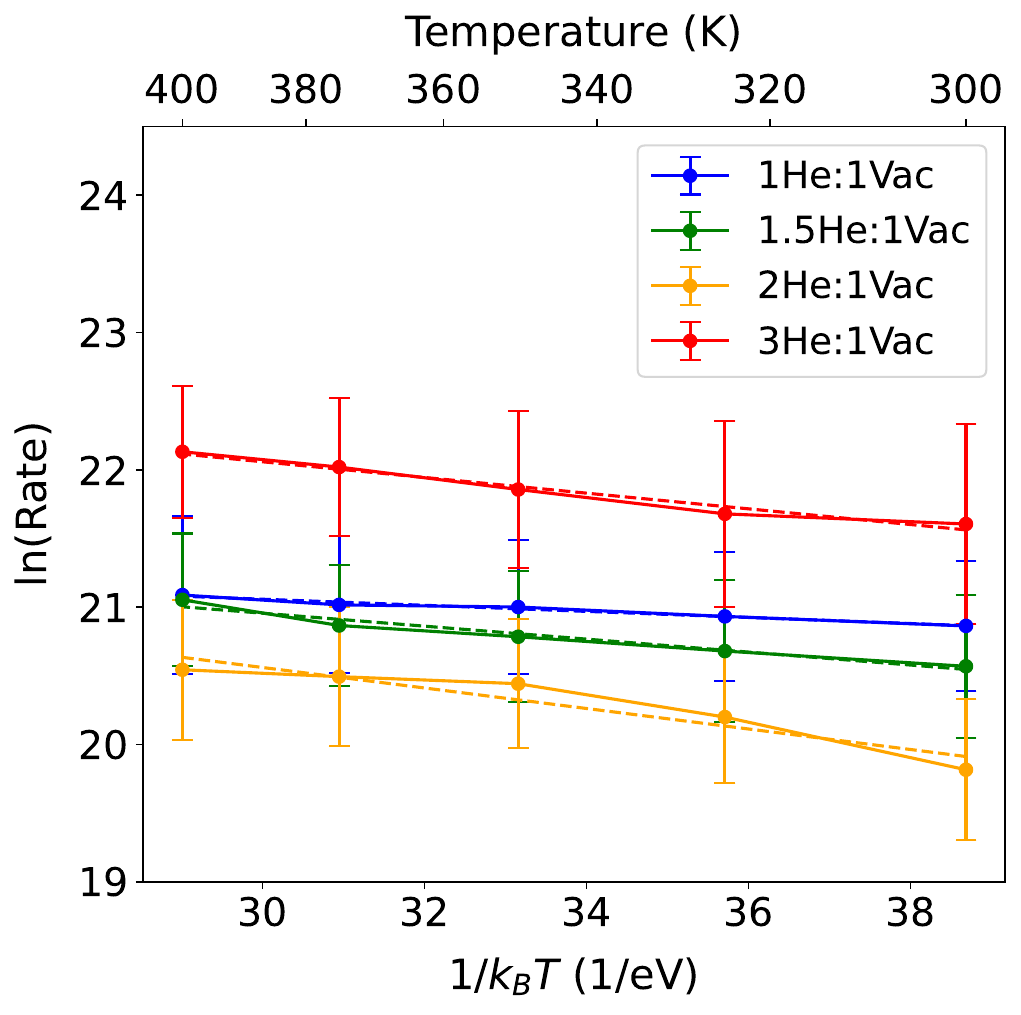} 
\caption{Dependence of the rate at which the dislocation overcomes He bubbles on temperature for bubbles of 1He:1Vac, 1.5He:1Vac, 2He:1Vac, and 3He:1Vac.
The rates are in units [ln(1/s)].
}
\label{fig:rate_vs_temperature}
\end{figure}

\begin{figure*}[!htbp]

\subfloat[\centering]{{\includegraphics[width=0.45\textwidth,trim=0.5cm 0.5cm 0.5cm 0.5cm, clip]{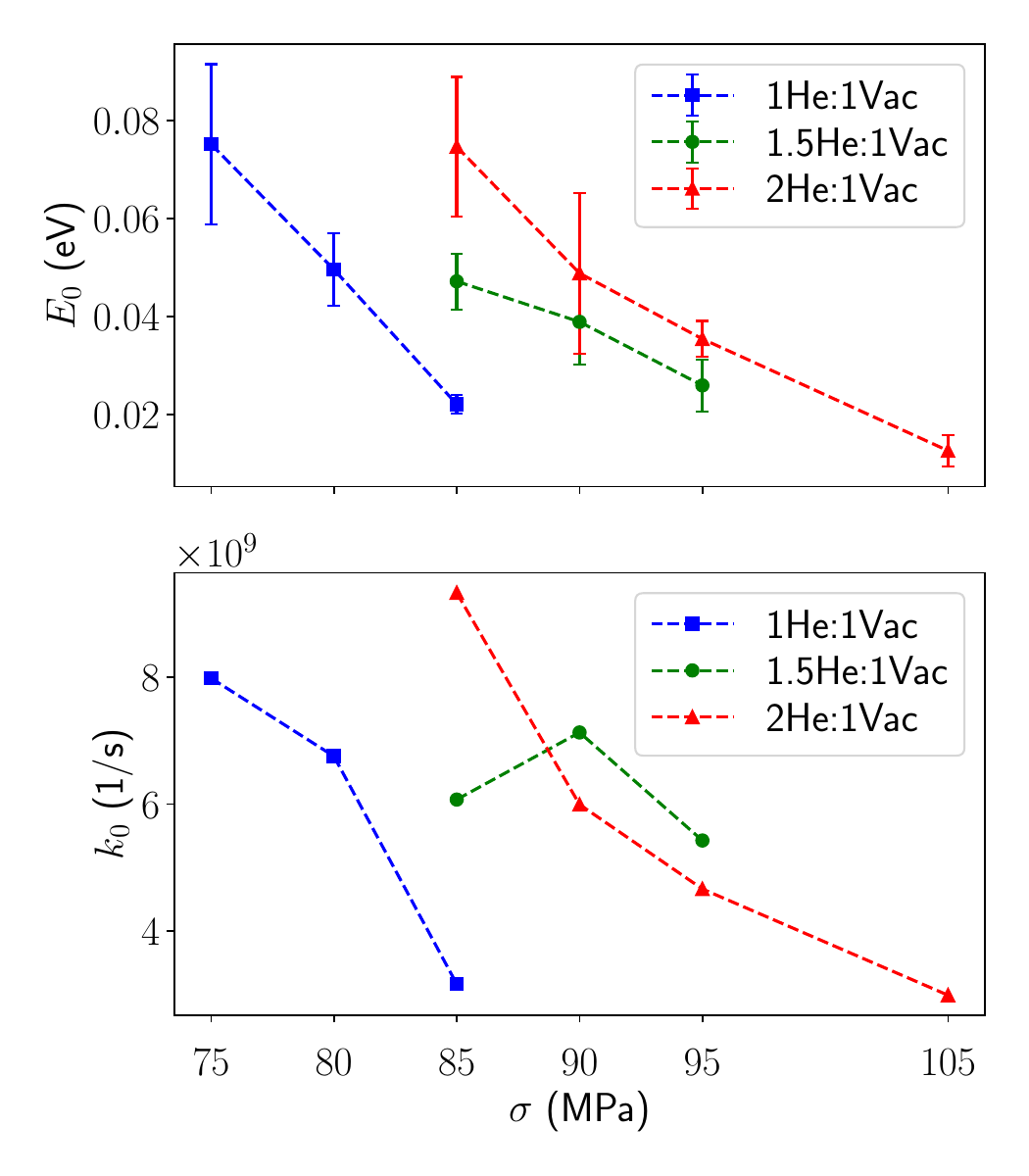}}}%
\qquad
\qquad
\subfloat[\centering]{{\includegraphics[width=0.45\textwidth]{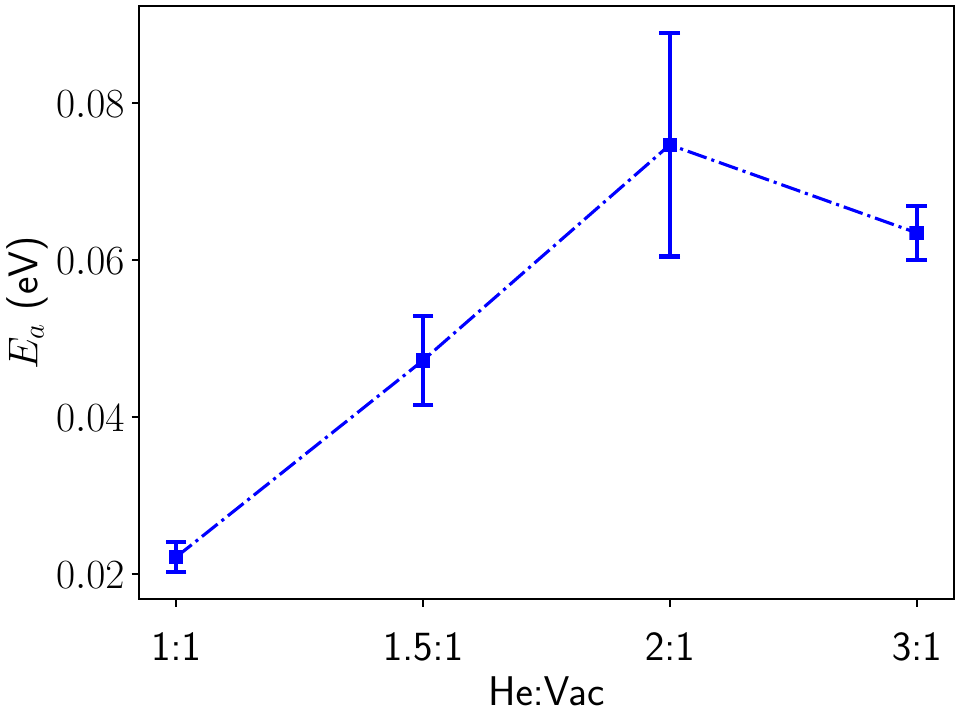} }}%
\caption{(a) Dependence of the activation energy and pre-factor on the applied shear stress for bubbles of 1He:1Vac, 1.5He:1Vac, and 2He:1Vac (b) The activation energy for He bubble-dislocation interactions at different He:Vac ratios under the applied stress of 85 MPa.}%
\label{fig:Ea_vs_stress}%
\end{figure*}
\begin{figure}[!htbp]
\includegraphics[width=\textwidth]{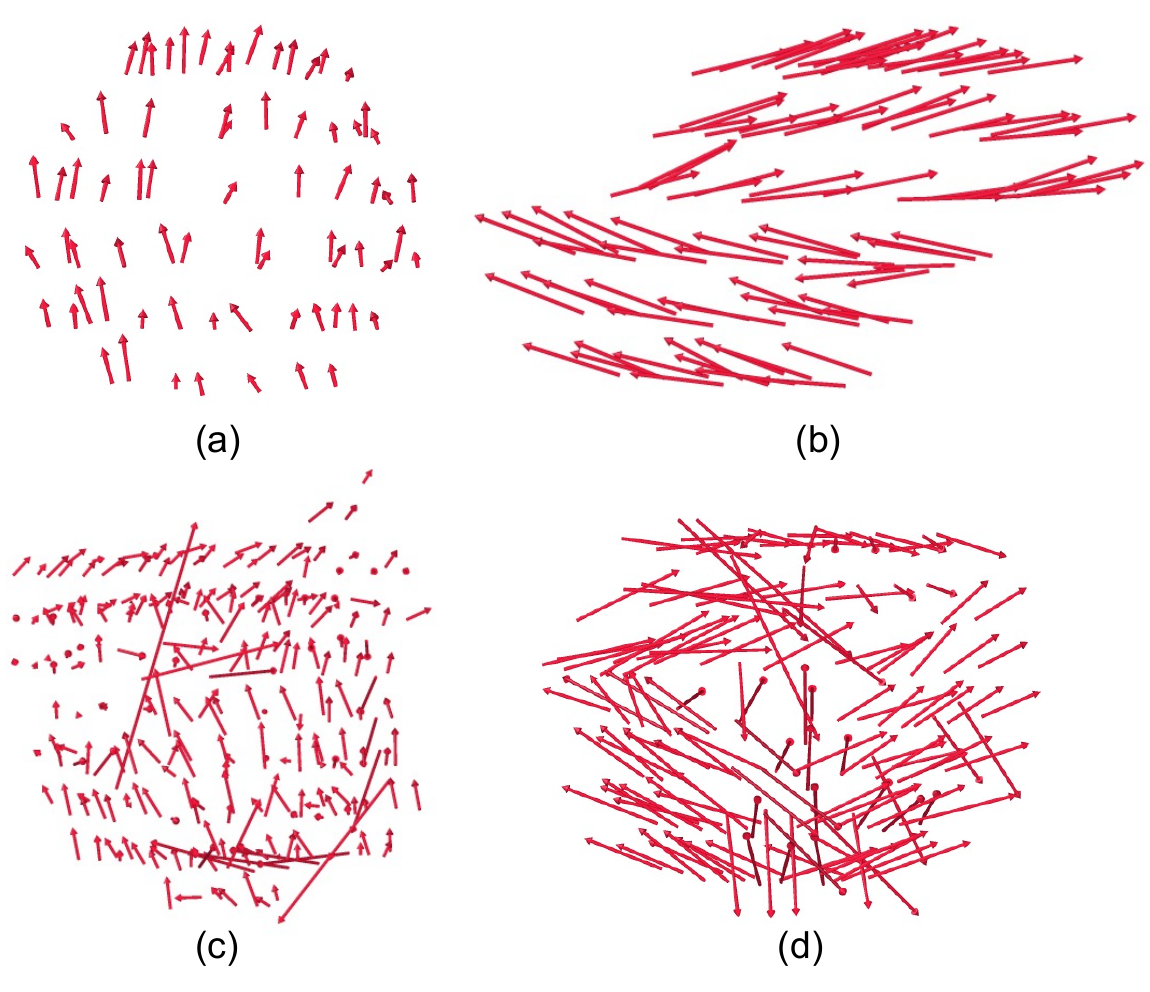}
\caption{The displacement vectors of He bubbles from (a) before the leading dislocation comes into contact with the 5\AA-radius bubble with 1.5He:1Vac (b) after the trailing dislocation leaves the 5\AA-radius bubble with 1.5He:1Vac.(c) before the leading dislocation comes into contact with the 5\AA-radius bubble with 3He:1Vac (d) after the dislocations leave the 5\AA-radius bubble with 3He:1Vac. The reference particle position is taken from the minimized configuration prior to shear loading.} 
\label{fig:shearing}
\end{figure}

To quantify the obstacle strength of the bubbles, we calculated the rate at which dislocations overcome He bubbles ($k$) at similar resolved shear stress (RSS) and temperature. Fig.~\ref{fig:rate_vs_temperature} shows $k$ at different temperatures and RSS of 85 MPa for different He:Vac ratios (1He:1Vac, 1.5He:1Vac, 2He:1Vac, 3He:1Vac). The rate for each condition is extracted as the average of six trials and the error bars shown in Fig.~\ref{fig:rate_vs_temperature} correspond to one standard deviation about the mean. As for bubbles of He:Vac up to 2:1, even though they demonstrate the same interaction mechanism, the interaction occurs at various rates at different He:Vac ratios. As shown in Fig.~\ref{fig:rate_vs_temperature}, He bubbles with higher He:Vac ratios are more resistant to the dislocation motion and require a greater amount of time to overcome them.

On the other hand, we notice that the rates at which a dislocation overcomes the bubble with 3He:1Vac are higher than those with smaller He:Vac ratios (He:Vac $\leq$ 2:1). This suggests that bubbles with 3He:1Vac, despite their higher concentration of He atoms, are weaker obstacles than those with 2He:1Vac (Fig.~\ref{fig:rate_vs_temperature}). The activation energies and pre-factors, estimated from the rates, are shown in Fig.~\ref{fig:Ea_vs_stress}
in order to rationalize the obstacle strength of He bubbles.
Fig.~\ref{fig:Ea_vs_stress}a shows the activation energy ($E_0$) and pre-factor ($k_0$) as a function of the applied shear stress for bubbles with 1He:1Vac, 1.5He:1Vac, and 2He:1Vac. 
As expected, the activation energy decreases with increase in RSS.
Note that the small values of activation energies are a consequence of the high RSS values we used in our study. Given the relatively large size of our simulations, the RSS values were chosen to achieve good statistics of dislocation bypass events in a reasonable amount of computational time.
The pre-factors are $\mathcal{O}({10^9} {s^{-1}})$, consistent with the vibration frequencies of the dislocation line, and also exhibit a weak dependence on the RSS. Fig.~\ref{fig:Ea_vs_stress}b compares the activation energy for dislocation interaction with He bubbles at different He:Vac under the same RSS of 85 MPa. It is evident that the activation energy increases with increase in He:Vac, reaches a peak at He:Vac $\leq$ 2:1, and then decreases. This indicates that there is a threshold helium-to-vacancy ratio at which the obstacle strength is maximized. Beyond this threshold, the bypass mechanism changes from bubble-shearing to super-jog formation and the strength drops.

Fig.~\ref{fig:shearing} demonstrates this change in mechanism. The arrows in this figure show the displacement vectors of He atoms in the bubble with respect to a reference configuration before applying the RSS. Panels (a) and (b) are for He:Vac =1.5, and in this case, the shearing is clearly visible in panel (b). This mechanism applies to He bubbles with He:Vac up to 2. However, due to the emission of Frenkel pairs, He bubbles with He:Vac =3 experience atomic re-arrangements in their structure and the shearing in panel (d) is less obvious compared to panel (b).
Highly-pressurized bubbles exhibit weaker obstacle strength compared to lower-pressure bubbles due to the emission of Frenkel pairs and subsequent formation of super-jogs. Ref. \cite{osetskystoller} reports a similar observation in BCC iron where the jog formation is observed for the 2 nm bubble with He:Vac =2 and this bubble demonstrates weaker obstacle strength compared to those with He:Vac $\leq$ 1:1. Also, previous studies in BCC iron point out that the formation of super-jogs contributes to the reduction in the critical applied shear stress for void-dislocation interactions \cite{osetsky_mohles} and He bubbles with high He:Vac ratios \cite{abe_he_iron}. Therefore, the conclusion that a bubble with higher He:Vac is a stronger obstacle to the mobility of dislocations only applies to the jog-free interaction mechanism.

\subsection{Interaction of moving edge dislocations with 10\AA-radius He bubbles}

\begin{figure*}[htb]
\includegraphics[width=\textwidth]{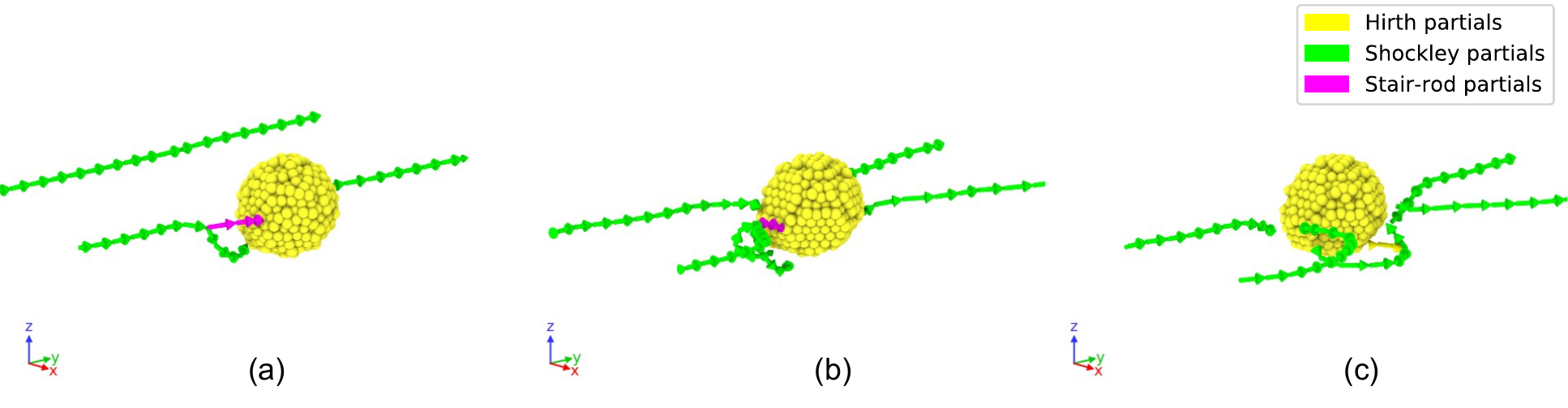}    
\caption{Interaction between the 10\AA-radius He bubble and dislocations. There are three stages considered in this interaction mechanism: (a) initial contact of the leading partial dislocation with the bubble (b) the leading and trailing partials are pinned simultaneously by the bubble (c) the partials are released from the bubble.
See also Table \ref{tab:dislocationlenngth_2X_10A}.} 
\label{fig:10A_DXA}
\end{figure*}

\begin{figure}[!htbp]

\includegraphics[width=\textwidth]{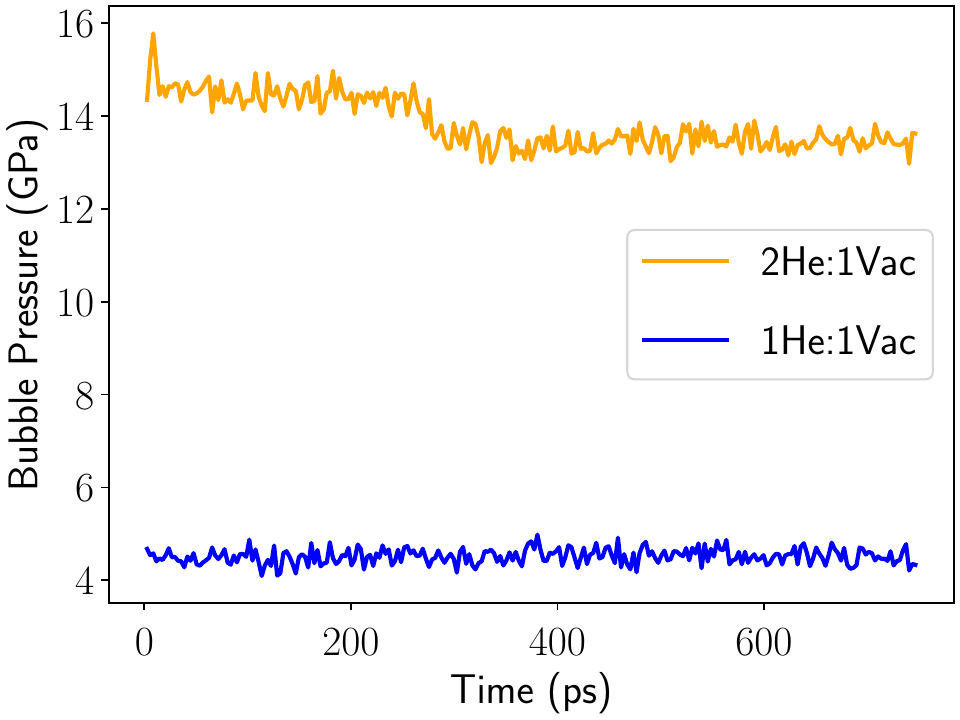}    
\caption{Changes in the bubble pressure during the interaction between dislocations and 10\AA-radius bubbles with 1He:1Vac and 2He:1Vac} 
\label{fig:10A_pressure}

\end{figure}

In order to understand the influence of bubble size on the dislocation motion, the interaction between moving dislocations and 10\AA-radius He bubbles was investigated. The 10\AA-radius bubble with He:Vac =1 exhibits a mechanism similar to the one described in the case of 5 \AA-radius bubbles with He:Vac $\leq$ 2:1. Fig.~\ref{fig:10A_DXA}, shows the interaction between the 10 \AA-radius bubble with He:Vac =2 and the dislocation. This mechanism is similar to the mechanism observed in the smaller bubble with He:Vac =3 (Fig.~\ref{fig:compare_dislocation_line_3x}.a). That is, after the leading partial comes into contact with the bubble, both partial dislocations are attached to it simultaneously. As a result, the partials constrict and a super-jog is formed as presented in Fig.~\ref{fig:10A_DXA}.c. In addition, the bubble pressure decreases by $\sim10\%$ ( Fig.~\ref{fig:10A_pressure}) indicating emission of Frenkel pairs and subsequent absorption of the vacancies by the bubble. The SIAs are absorbed by the dislocation leading to the formation of a super-jog.  Although this mechanism is similar to the 5 \AA-radius bubble with He:Vac =3, it is interesting that the transition from shearing to a jog-mediated mechanism happens at a lower initial pressure of $\sim14$ GPa for the 10 \AA-radius bubble, compared to $\sim23$ GPa for the 5 \AA-radius bubble.
This shift in the `critical' He:Vac ratio with size of the He bubble was also reported previously for simulations in Fe \cite{osetskystoller}.
This suggests that the transition is not solely driven by bubble pressure.
An influencing factor could be the atomic arrangement in the bubble. Previous work shows that helium atoms in bubbles within iron can form an ordered structure, which is likely body-centered cubic (bcc) at low helium contents and hexagonal close-packed (hcp) at high helium contents \cite{morishita2003thermal}. The degree of coherence between the helium bubble configuration and the metal matrix can significantly affect the formation of self-interstitial atoms (SIAs) and their emission, ultimately contributing to jog formation. To investigate this hypothesis, we quantified the structural transitions in our simulations as described below.

\begin{table}[!htbp]

\caption{Dislocation length in the micro-structure of the system during the interaction between dislocations and the 10\AA-radius He bubble.
The three stages correspond to those presented in Fig.~\ref{fig:10A_DXA}.}
\begin{tabular}{|p{1cm}||p{1.75cm}|p{1.75cm}|p{1.75cm}|}

 \hline
 \multicolumn{4}{|c|}{He:Vac = 2:1} \\
 \hline
 Stage & Shockley partial (nm) & Stair-rod partial (nm) & Hirth partial (nm)\\
 \hline
 1   & 21.7    & 0.12&   0\\
 2&   22.6 & 0.89   & 0\\
 3 &23.1 & 0 &  0.14\\
 
 \hline
\end{tabular}
\label{tab:dislocationlenngth_2X_10A}
\end{table}

\begin{figure}
\includegraphics[width=\textwidth]{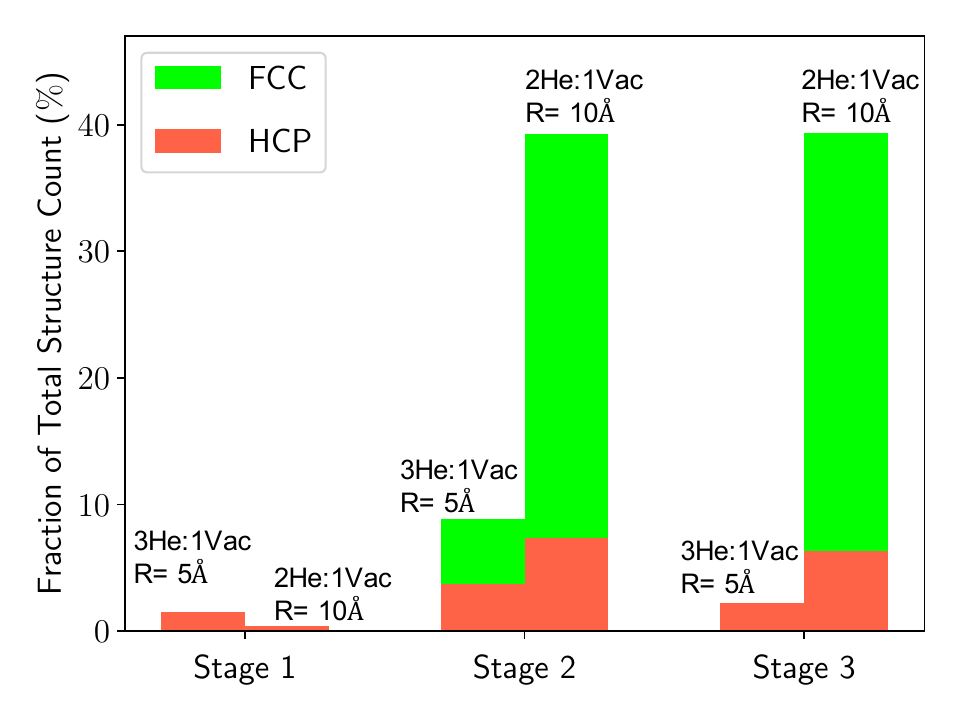}    
\caption{Distribution of He atom arrangements obtained from adaptive-common neighbor analysis. Due to the similarity in the interaction mechanism, 5\AA-radius bubble with 3He:1Vac and 10\AA-radius bubble with 2He:1Vac are selected. Only atoms of FCC and HCP types are presented in this figure, the remaining fraction is counted for disordered atoms. Three stages correspond to those presented in Fig.~\ref{fig:10A_DXA}.} 
\label{fig:StructureCount}
\end{figure}

Our simulations predict differences in the atomic arrangement of He atoms and the distribution of partial dislocations between the 10\AA-radius (2He:1Vac) and 5\AA-radius bubbles (3He:1Vac). As shown in Fig.~\ref{fig:StructureCount}, when the leading partial comes into contact  with the bubble (stage 1), the majority of He atoms are characterized as disordered atoms using CNA. Less than 5\% are characterized as hexagonal close packed (HCP) atoms in 5\AA-radius bubbles whereas a mere fraction is identified as HCP atoms in the 10\AA-radius bubbles. At stages 2 and 3, the fraction of face-centered cubic (FCC) and HCP atoms reaches approximately 35\% and 6-7\%, respectively, for the large bubble. This indicates that the interaction with the leading partial leads to ordering of the He atoms in the larger bubble and this ordering persists as the trailing partial cuts through. Meanwhile, the fractions of FCC and HCP atoms in the smaller bubble are approximately 8\% and 3\%, respectively, at stage 2 indicating no such ordering.
We note that the CRSS for the dislocation to overcome 10\AA-radius (2He:1Vac) bubbles is $\sim 140$ MPa, clearly demonstrating that they are stronger obstacles compared to the smaller bubbles and suggest a possible explanation for the shift in the `critical' He:Vac ratio.

\section{Conclusions}
In this work, we have investigated the interaction between dislocations and He bubbles in FCC Cu using molecular dynamics simulations. 
He bubbles with 1 nm and 2 nm  diameter and He:Vac ratios ranging from 1 to 3 are the main focus in this work.
The main results are summarized as follows:

The mechanism by which the dislocation overcomes the He bubble obstacle is dependent on the He:Vac ratio of the bubble. Bubbles with smaller He:Vac are sheared while a jog-mediated mechanism is predicted at higher He:Vac.
For bubbles that are sheared, the obstacle strength increases with the increase in He:Vac. Bubbles which are overcome through a jog-mediated mechanism are weaker obstacles compared to those that are sheared.
The transition from shearing to a jog-mediated mechanism is a function on both bubble size and He:Vac ratio.
Larger bubbles are found to transition to the the jog-mediated mechanism at lower He:Vac (or equivalently, low pressure).
Ordering of He atoms is predicted in larger bubbles during the interaction with the dislocation. We hypothesize that this ordering could play a role in the higher obstacle strength of larger bubbles. 
Activation energies and pre-factors obtained for the dislocation-bubble interaction could be used to develop higher-length scale models for this process.

Future studies on the interaction between helium bubbles and dislocations in fcc-Cu can be extended to high-temperature environments.
At low temperatures, helium bubbles cause significant hardening and embrittlement due to their strong pinning effect on dislocations. However, at higher temperatures, this effect is mitigated as dislocations can more easily bypass the bubbles.
Elevated temperatures can also enhance atomic diffusion, allowing for increased movement of point defects, such as interstitial atoms, which can have a significant effect on dislocation-bubble interaction.
Therefore, understanding the interaction between helium bubbles and dislocations across a wide range of temperatures can provide valuable insights into how macroscopic properties such as strength, ductility, toughness, creep, and overall structural integrity are affected.
This is left for future work.

\section{Acknowledgement}

The authors are grateful for the support of the Materials project within the Advanced Simulation and Computing, Physics and Engineering Models Program of the U.S. Department of Energy under contract 89233218CNA000001.

\setlength{\parindent}{0pt}
\setlength{\bibsep}{4pt}
\setlength\itemindent{0pt}
\setlength{\bibhang}{\bibmargin}
{\setstretch{0.85}
\fontsize{8}{10} \selectfont
\bibliographystyle{IEEETran_custom}
\renewcommand\refname{References}  
\bibliography{References}}
\end{document}